%% file: main.tex
\documentclass[conference,comsoc]{IEEEtran}
\IEEEoverridecommandlockouts
\usepackage{cite}
\usepackage{amsmath,amssymb,amsfonts}
\usepackage{algorithmic}
\usepackage{graphicx}
\usepackage{textcomp}

\usepackage{float}
\usepackage{graphics}
\usepackage{kvoptions}
\usepackage{tikz}
\usepackage{titlecaps}
\usepackage{hyperref}

\newcommand\copyrighttext{%
  \footnotesize \textcopyright 2017 IEEE. Personal use of this material is permitted.
  Permission from IEEE must be obtained for all other uses, in any current or future
  media, including reprinting/republishing this material for advertising or promotional
  purposes, creating new collective works, for resale or redistribution to servers or
  lists, or reuse of any copyrighted component of this work in other works.
  DOI: \href{https://doi.org/10.1109/CIG.2017.8080450}{10.1109/CIG.2017.8080450}}
\newcommand\copyrightnotice{%
\begin{tikzpicture}[remember picture,overlay]
\node[anchor=south,yshift=10pt] at (current page.south) {\fbox{\parbox{\dimexpr\textwidth-\fboxsep-\fboxrule\relax}{\copyrighttext}}};
\end{tikzpicture}%
}

\font\myfont=cmr12 at 22pt
\title{\myfont Cellular Automata Simulation on FPGA for Training Neural Networks with Virtual World Imagery}
\author{
  \IEEEauthorblockN{\makebox[.5\linewidth]{Olivier Van Acker and Oded Lachish}}
  \IEEEauthorblockA{Department of Computer Science and Information Systems\\
      Birkbeck, University of London\\ 
      London, United Kingdom\\
      olivier@dcs.bbk.ac.uk, oded@dcs.bbk.ac.uk} 
  \and
  \IEEEauthorblockN{\makebox[.5\linewidth]{Graeme Burnett}}
  \IEEEauthorblockA{Enhyper Ltd.\\ 
      London, United Kingdom\\
      graeme.burnett@ieee.org}
}
\begin{document}
\maketitle
\copyrightnotice
\begin{abstract}
We present ongoing work on a tool that consists of two parts: (i) A raw micro-level abstract world simulator with an interface to (ii) a 3D game engine, translator of raw abstract simulator data to photorealistic graphics. Part (i) implements a dedicated cellular automata (CA) on reconfigurable hardware (FPGA) and part (ii) interfaces with a deep learning framework for training neural networks. The bottleneck of such an architecture usually lies in the fact that transferring the state of the whole CA significantly slows down the simulation. We bypass this by sending only a small subset of the general state, which we call a 'locus of visibility', akin to a torchlight in a darkened 3D space, into the simulation. The torchlight concept exists in many games but these games generally only simulate what is in or near the locus. Our chosen architecture will enable us to simulate on a micro level outside the locus. This will give us the advantage of being able to create a larger and more fine-grained simulation which can be used to train neural networks for use in games.
\end{abstract}


\section{Background}
There have been exciting new results of training neural networks with photo realistic imagery from virtual worlds \cite{noauthor_artificial_2017}. 
The training of these neural networks uses rendered images from virtual worlds instead of real world data, the two biggest advantages of this approach being, firstly, fewer limitations in executing potentially difficult or dangerous scenarios, and secondly, the ability to accelerate the speed of the simulation means faster training of the neural network.

In recent years, the game industry has spent a lot of effort on creating game engines which can output near-photorealistic imagery in real time, making it possible to train neural networks for real world scenarios using this output. Several projects are being developed to make it easier for neural network frameworks to interface with these engines \cite{qiu_unrealcv:_2016}, \cite{kinsley_h_explorations_2017}.

Cellular Automata (CA) is an effective technique for simulating, on a micro level, complex behavior such as pedestrian traffic, moving agents \cite{blue_cellular_2001} 
or, as in our proof of concept, the traffic of narrowboats\footnote{Narrowboats were the main transportation system for goods at the start of the industrial revolution in the UK} on a system of canals. With a simple set of rules, contained in each cell, describing the behavior of passing agents, it is possible to get an extraordinarily complex macroscopic view of the flow of traffic \cite{blue_cellular_2001}.

CA, because of its inherent massive spatial parallelism, locality and discrete nature, is a perfect candidate for implementation on programmable FPGA (field-programmable gate array) technology \cite{halbach_implementing_2004}. FPGAs are reconfigurable hardware devices, where the set of rules contained in each cell can be described in hardware via lookup tables and flip-flops (for storing state), and with every clock cycle the state of all cells can be updated in parallel.

The remainder of this paper describes our current work, in which we are implementing a CA for simulating traffic on an FPGS, to train neural networks. These trained neural networks can be used to create game environments with 'real world'-like behavior.

\section{Implementation of ca Microsimulation on fpga}
We are developing a tool which microsimulates traffic in a virtual world and gives a game engine a limited view of certain areas of the world – a 'locus of visibility'. By limiting the amount of data made available for rendering and subsequent learning of a neural network, we can increase the size and granularity of the simulation, which will make the macro view more realistic.

The data of this locus of view, akin to a torchlight in a darkened 3D space, or a traffic camera used to monitor traffic on a busy crossing in a city, will be passed on to a game engine which will generate a photorealistic video feed of the exposed area. This video feed will be used for training a neural network. The simulation will run several times faster than it would normally do when a game is played, to speed up the training of the neural network.

The simulation uses the cellular automata (CA) method and a 'locus of visibility controller' extracts the localized data from the simulation and exposes it over the network to one or more consumers.

The CA will be implemented on a FPGA and communicate over PCIe to a network card (NIC), exposing it over the network. The Unreal Engine captures this data and a neural network interfaces with the game engine via UnrealCV \cite{qiu_unrealcv:_2016}, an\newpage 
\begin{figure}[H]
  \centering
  \def\svgwidth{1.0\linewidth}
  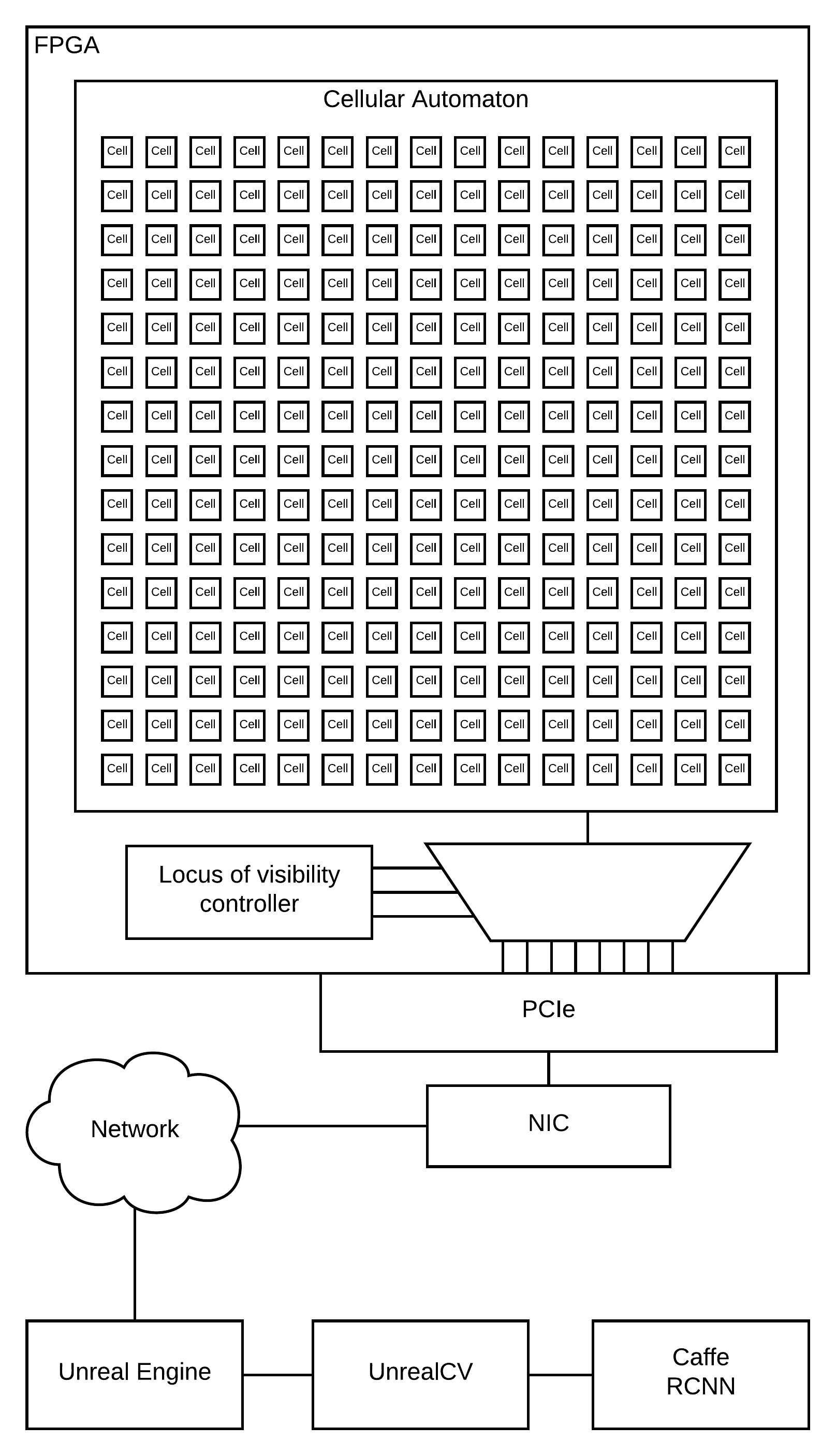
  \caption[Tool Architecture]{Architecture of the tool}
  \label{fig:fpga_ca}
\end{figure}
Unreal Engine plugin for interfacing with deep learning frameworks such as Caffe \cite{jia_caffe:_2014}.

The simulated world is split up into hexagons and each individual area has a fixed locality to adjacent hexagons and can hold multiple agents. \mbox{Each hexagon is represented by a cell} in the CA. The movement of agents and the state of the cell is determined by the rule set contained in every individual cell.
\newpage
\subsection{Proof of concept: Narrowboat Simulator}
In our first proof of concept we will simulate the traffic of a large number of narrowboats on an intricate canal system, transporting resources (for example, coal or grain) between supply points and delivery points in competing economic areas. The flow and density of traffic can be regulated via locks (chambered gates used to raise or lower water level, allowing boats to move to higher or lower levels of a canal), distributed throughout the canal system.

The neural network first trains itself by observing the traffic at several locks. Once trained, the neural network will be used to play a game in which it competes against a (human) opponent, to bring as many resources as possible to its own delivery points, manipulating the traffic by operating the locks.
\section*{Future Work}
The implementation of the narrowboat simulator will be a starting point from which to build more complex simulations of growing cities with different transportation systems, interacting with each other. This city transport simulator can then be used to train neural networks to operate different aspects of the simulation, for example, resource management. Both trained neural networks and simulator can be used to create games in which the environments and elements within them exhibit more complex, 'real world'-like behavior.

\bibliography{Zotero}{}
\bibliographystyle{IEEEtran}

\end{document}

%% file: Architecture.pdf_tex
\begingroup%
  \makeatletter%
  \providecommand\color[2][]{%
    \errmessage{(Inkscape) Color is used for the text in Inkscape, but the package 'color.sty' is not loaded}%
    \renewcommand\color[2][]{}%
  }%
  \providecommand\transparent[1]{%
    \errmessage{(Inkscape) Transparency is used (non-zero) for the text in Inkscape, but the package 'transparent.sty' is not loaded}%
    \renewcommand\transparent[1]{}%
  }%
  \providecommand\rotatebox[2]{#2}%
  \ifx\svgwidth\undefined%
    \setlength{\unitlength}{465bp}%
    \ifx\svgscale\undefined%
      \relax%
    \else%
      \setlength{\unitlength}{\unitlength * \real{\svgscale}}%
    \fi%
  \else%
    \setlength{\unitlength}{\svgwidth}%
  \fi%
  \global\let\svgwidth\undefined%
  \global\let\svgscale\undefined%
  \makeatother%
  \begin{picture}(1,1.74193548)%
    \put(0,0){\includegraphics[width=\unitlength,page=1]{Architecture.pdf}}%
  \end{picture}%
\endgroup%

%% file: main.bbl
\begin{thebibliography}{1}
\providecommand{\url}[1]{#1}
\csname url@samestyle\endcsname
\providecommand{\newblock}{\relax}
\providecommand{\bibinfo}[2]{#2}
\providecommand{\BIBentrySTDinterwordspacing}{\spaceskip=0pt\relax}
\providecommand{\BIBentryALTinterwordstretchfactor}{4}
\providecommand{\BIBentryALTinterwordspacing}{\spaceskip=\fontdimen2\font plus
\BIBentryALTinterwordstretchfactor\fontdimen3\font minus
  \fontdimen4\font\relax}
\providecommand{\BIBforeignlanguage}[2]{{%
\expandafter\ifx\csname l@#1\endcsname\relax
\typeout{** WARNING: IEEEtran.bst: No hyphenation pattern has been}%
\typeout{** loaded for the language `#1'. Using the pattern for}%
\typeout{** the default language instead.}%
\else
\language=\csname l@#1\endcsname
\fi
#2}}
\providecommand{\BIBdecl}{\relax}
\BIBdecl

\bibitem{noauthor_artificial_2017}
\BIBentryALTinterwordspacing
``Artificial intelligence: {Why} {AI} researchers like video games {\textbar}
  {The} {Economist},”,'' May 2017. [Online]. Available:
  \url{http://www.economist.com/news/science-and-technology/21721890-games-help-them-understand-reality-why-ai-researchers-video-games}
\BIBentrySTDinterwordspacing

\bibitem{qiu_unrealcv:_2016}
\BIBentryALTinterwordspacing
W.~Qiu and A.~Yuille, ``{UnrealCV}: {Connecting} {Computer} {Vision} to
  {Unreal} {Engine},'' \emph{arXiv:1609.01326 [cs]}, Sep. 2016, arXiv:
  1609.01326. [Online]. Available: \url{http://arxiv.org/abs/1609.01326}
\BIBentrySTDinterwordspacing

\bibitem{kinsley_h_explorations_2017}
H.~Kinsley, ``Explorations of {Using} {Python} to play {Grand} {Theft} {Auto}
  5.'' 2017.

\bibitem{blue_cellular_2001}
\BIBentryALTinterwordspacing
V.~J. Blue and J.~L. Adler, ``Cellular automata microsimulation for modeling
  bi-directional pedestrian walkways,'' \emph{Transportation Research Part B:
  Methodological}, vol.~35, no.~3, pp. 293--312, Mar. 2001. [Online].
  Available:
  \url{http://www.sciencedirect.com/science/article/pii/S0191261599000521}
\BIBentrySTDinterwordspacing

\bibitem{halbach_implementing_2004}
M.~Halbach and R.~Hoffmann, ``Implementing cellular automata in {FPGA} logic,''
  in \emph{18th {International} {Parallel} and {Distributed} {Processing}
  {Symposium}, 2004. {Proceedings}.}, Apr. 2004, pp. 258--.

\bibitem{jia_caffe:_2014}
\BIBentryALTinterwordspacing
Y.~Jia, E.~Shelhamer, J.~Donahue, S.~Karayev, J.~Long, R.~Girshick,
  S.~Guadarrama, and T.~Darrell, ``Caffe: {Convolutional} {Architecture} for
  {Fast} {Feature} {Embedding},'' in \emph{Proceedings of the 22Nd {ACM}
  {International} {Conference} on {Multimedia}}, ser. {MM} '14.\hskip 1em plus
  0.5em minus 0.4em\relax New York, NY, USA: ACM, 2014, pp. 675--678. [Online].
  Available: \url{http://doi.acm.org/10.1145/2647868.2654889}
\BIBentrySTDinterwordspacing

\end{thebibliography}
